\documentclass[10pt,pra,twocolumn,superscriptaddress,showpacs,floatfix]{revtex4-2}
\usepackage{graphics}
\usepackage{graphicx}
\usepackage{bm}
\usepackage{amsmath}
\usepackage{amsfonts}
\usepackage{amssymb}
\usepackage{latexsym}
\usepackage{soul}
\usepackage[dvipsnames]{xcolor}
\usepackage{hyperref}
\usepackage{float}

\begin{document}

\title[Enhancing wave-particle duality]{Enhancing wave-particle duality}

\author{Arwa Bukhari}
\address{School of Physics and Astronomy, University of Leeds, Leeds LS2 9JT, United Kingdom}
\address{School of Physics, Umm Al-Qura University, Makkah, Saudi Arabia}
\author{Daniel Hodgson}
\address{School of Physics and Astronomy, University of Leeds, Leeds LS2 9JT, United Kingdom}
\author{Sara Kanzi}
\address{Faculty of Engineering, Final International University, North Cyprus Via Mersin 10, Kyrenia, 99370, Turkey}
\author{Robert Purdy}
\address{School of Physics and Astronomy, University of Leeds, Leeds LS2 9JT, United Kingdom}
\author{Almut Beige}
\address{School of Physics and Astronomy, University of Leeds, Leeds LS2 9JT, United Kingdom}
\vspace{10pt}

\begin{abstract}
To enhance the consistency between the quantum descriptions of waves and particles, we quantise mechanical point particles in this paper in the same physically-motivated way as we previously quantised light in quantum electrodynamics [Bennett {\em et al.}, Eur.~J.~Phys.~{\bf 37}, 014001 (2016)]. To identify the relevant Hilbert space, we notice that mechanical particles can occupy any position $x$ while moving at any velocity $v$. Afterwards, we promote the classical states $(x,v)$ to pairwise orthogonal quantum states $|x,v\rangle$ and demand that these evolve according to Newton's equations of motion. The resulting quantum theory is mass-independent, when Newton's equations of motion are mass-independent, as one would expect. The basic formulation of quantum mechanics emerges from quantum mechanics in configuration space as a semi-classical approximation when a fixed mass is imposed and several other adjustments are made.
\end{abstract}

%
%
%
\maketitle
%
%

\section{Introduction} \label{sec1}

In classical physics, Hamilton's formulation of the laws of mechanics specifies the motion of a point particle moving in one-dimension in terms of two quantities: its position $x$ and its momentum $p$ \cite{Griffith}. The dynamics of the position $x$ and the momentum $p$ can be found by solving Hamilton's equations of motion \cite{Dirac,Goldstein},
\begin{eqnarray} \label{Hamilton1}
\frac{\partial x}{\partial t} = \frac{\partial H}{\partial p} \, ,  ~~ 
\frac{\partial p}{\partial t} = -\frac{\partial H}{\partial x} \, ,
\end{eqnarray}
where $H=H(x,p)$ is the relevant Hamiltonian. When solving these equations, we obtain a function $x(t)$ which defines the trajectory of the particle. As pointed out by Dirac \cite{Dirac2}, if the Hamiltonian $H$ of the system is not appropriately defined, Eq.~(\ref{Hamilton1}) can lead directly to inconsistencies in the theory. Nevertheless, Hamiltonian mechanics lends itself naturally to the quantisation of mechanical point particles. In canonical quantisation, this is done by identifying the observables for the position and the momentum of quantum mechanical point particles with a pair of canonically conjugate operators which we denote $\hat{x}$ and $\hat{p}$ \cite{Hi,Schr,Von,Rae,Liboff,Englert,Franz,MQM}. 

Notice that Hamiltonian mechanics provides just one formulation of classical mechanics, and other alternative descriptions are available. One example is the earlier Newtonian description \cite{MQM,Goldstein}, which we adopt in the following as the starting point for a physically-motivated quantisation of mechanical point particles that move in the presence of a non-zero potential $V(x)$ along the $x$ axis. Newton's second law states that the force $F(x) = - \partial V(x)/ \partial x$ experienced by a particle at position $x$ equals
\begin{eqnarray} \label{1}
F(x) &=& m \, \ddot x
\end{eqnarray}
where $m$ and $\ddot x= \partial^2 x/\partial t^2$ denote its rest mass and its acceleration respectively. To turn the second order differential equation (\ref{1}) into two first order differential equations, Newton, and also Leibniz, described mechanical particles by specifying not only their position $x$ but also their velocity $v$. The result is Newton's equations of motion,
\begin{eqnarray} \label{Newton4}
\frac{\partial x}{\partial t} = v \, , ~~ \frac{\partial v}{\partial t} = f(x) 
\end{eqnarray}
with the function $f(x)$ representing a position-dependent force per mass, 
\begin{eqnarray}\label{newsch2}
f(x) &=& - {1 \over m} \, \frac{\partial V(x)}{ \partial x} \, .
\end{eqnarray}
This variable is the key variable which fully determines the dynamics of a mechanical point particle in a potential $V(x)$ with given initial conditions, like its initial position and velocity, $x(0)$ and $v(0)$.

The starting point for the quantisation of mechanical point particles in the basic formulation of quantum mechanics (QM) which we review in Section \ref{sec2} is Hamilton's formalism. In this paper, we quantise instead Newton's equations of motion while taking the whole configuration space and not only a subspace of states of mechanical point particles into account. The term {\em configuration space} is the fundamental space of all possible states that a physical system can be in. As we shall see below, the resulting formalism, which we refer to as {\em QM in configuration space}, has several advantages. For example, its Schr\"odinger equation is mass independent, if the same is true for Newton's equations of motion in Eq.~(\ref{Newton4}). This connection increases the consistency between quantum and classical mechanics. Moreover, QM in configuration space enhances wave-particle duality, i.e.~the consistency between the quantum descriptions of wave and particles, by quantising mechanical point particles in the same physically-motivated way as we quantise the electromagnetic field and its photonic carriers in free space \cite{Bennett,Jake,Daniel,AMC}.

The first step of physically-motivated quantisation schemes is to identify the relevant configuration space. Afterwards, we associate its classical states with pairwise orthonormal quantum states. These are the {\em most classical quantum states} of the physical system and form a complete basis of its Hilbert space ${\cal H}$. In addition, we assume in the following that all quantum states evolve according to the Schr\"odinger equation based on a {\em dynamical Hamiltonian} $\hat H_{\rm dyn}$. This Hamiltonian can be derived by demanding that the most classical quantum states evolve in a way that is consistent with classical physics. As we have seen already in the case of photons, $\hat H_{\rm dyn}$ is in general unbounded from below \cite{Jake,Daniel}. It therefore does not coincide with the energy observable of the quantum system, although there is a close connection between the energy observable and $H_{\rm dyn}$ \cite{AMC}. Notice also that our quantisation approach \cite{Bennett} has some similarities with the quantum reconstruction program \cite{Ludwig,Hardy} which already attracted some attention \cite{Grinbaum,Spekkens,Masanes,Berghofer}.

As mentioned previously, in Newtonian mechanics we describe the dynamics of a mechanical point particle by specifying its position $x$ and its velocity $v$ at all times $t$. In fact, the initial position $x(0)$ and the initial velocity $v(0)$ of the particle can be chosen independently. To ensure that position and velocity are of equal importance in the corresponding quantum theory, we promote the distinguishable classical states $(x,v)$ to orthonormal quantum states $|x,v \rangle$ with
\begin{eqnarray} \label{delta1}
\langle x,v| x',v' \rangle &= \delta(x-x') \, \delta(v-v') \, . 
\end{eqnarray}
These states form a basis in the Hilbert space ${\cal H}$, i.e.~the configuration space, of quantum mechanical point particles. Although the $|x,v \rangle$ states have some similarities with the coherent states $|z \rangle = |x + {\rm i}p \rangle$ that represent the classical phase space in alternative formulations of QM \cite{Dorje,Caslav,Gazeau,PS,PS2,PS3,Rundle,Matinez}), they are not the same. While the $|x,v \rangle$ states describe individual particles, the $|z \rangle$ states are many-body quantum states. An important difference between basic QM and QM in configuration space is that the Hilbert space ${\cal H}$ of the latter is significantly larger and allows quantum mechanical point particles to be simultaneously localised in position {\em and} in velocity. As we shall see below, the momentum and the velocity of a particle represent two independent physical observables. In general, the quantum state $|\psi(t) \rangle $ of a quantum mechanical point particle is a superposition of the $|x,v \rangle$ states with $|\langle \psi(t) | x,v \rangle|^2$ representing the probability for finding the particle at time $t$ at position $x$ while moving at velocity $v$. 

\begin{figure}[t]
\centering
\includegraphics[width=0.99 \linewidth] {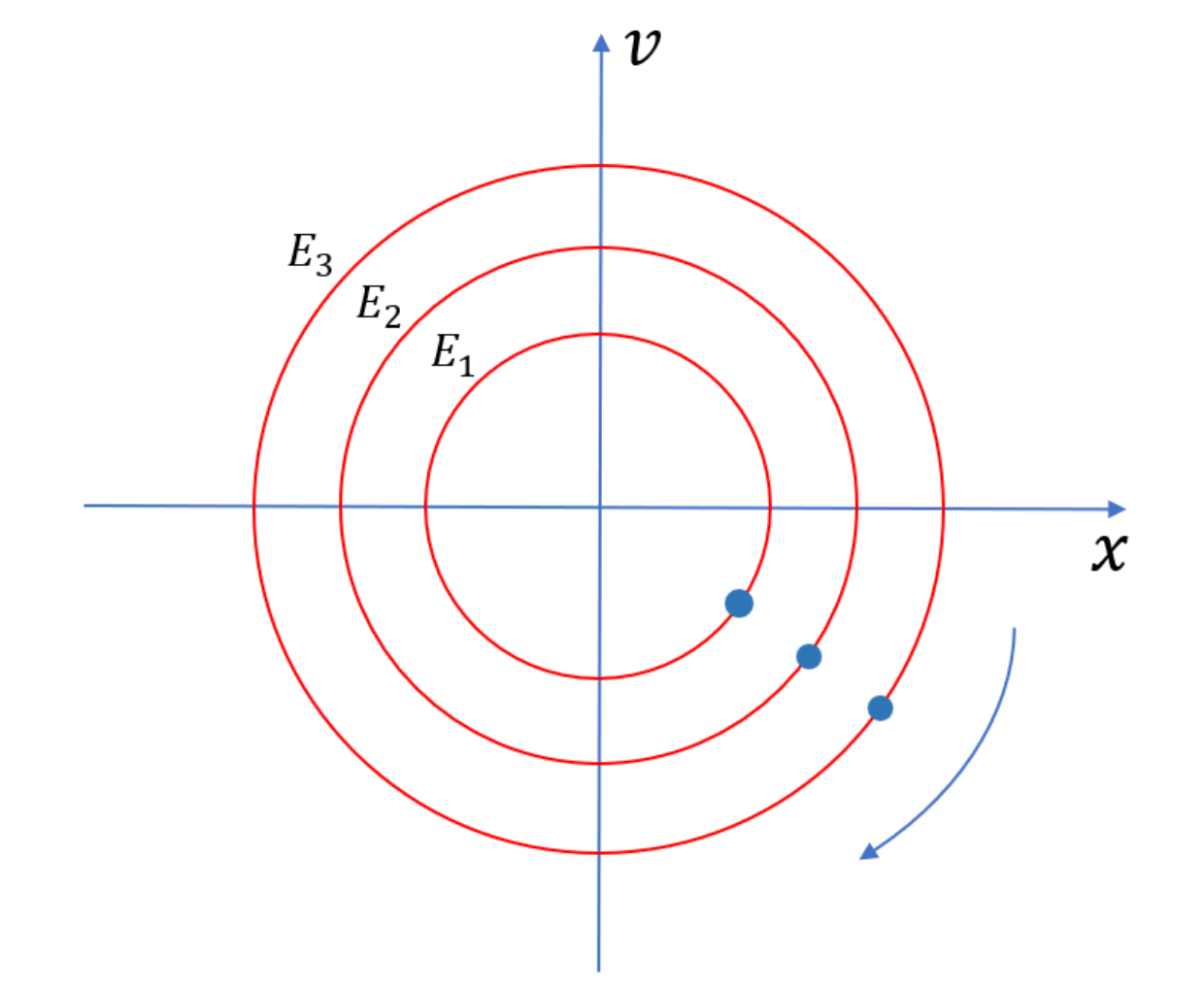}
\caption{The classical phase space trajectories (red) of three mechanical point particles with energies $E_n$ $(n=1,2,3)$ moving along positions $x_n(t)$ with velocities $v_n(t)$ inside a harmonic potential. As time progresses, these trajectories form closed circles. The radius of each circle depends on the energy $E_n$ of the $n$-th particle with the more energetic particles being further away from the origin of the phase space diagram. In the following, we demand that the ``most classical" quantum states $|x(t),v(t) \rangle$ of quantum mechanical point particles (represented by the blue dots) remain localised at all times and evolve along the corresponding classical trajectory. In this way, any classical experiment with a point particle inside a harmonic oscillator can also be described by quantum physics.}
\label{fig:c}
\end{figure}

Since the state $|x,v \rangle$ represents the $(x,v)$ state of a classical point particle in the corresponding quantum theory, consistency between classical mechanics and QM in configuration space requires that this state remains localised at all times.  In addition, its position $x$ and its velocity $v$ must evolve according to Newton's equations of motion. In the following, we identify the dynamical Hamiltonian $\hat H_{\rm dyn}$ that generates the above dynamics and ensures that the variables $x(t)$ and $v(t)$ follow their classical trajectories in configuration space (cf.~e.g.~Fig.~\ref{fig:c}). Once the dynamical Hamiltonian $\hat H_{\rm dyn}$ is known, we also know how all other possible quantum states $|\psi(t) \rangle$  evolve in time.

Since we quantise mechanical point particles in this paper in the same physically motivated way as we quantise light in quantum electrodynamics, QM in configuration space enhances wave-particle duality \cite{Fano} which plays an important role for quantum technology applications \cite{Janovitch}. It can be used to model both the dynamics of photons and the dynamics of non-relativistic quantum mechanical point particles equivalently. The crucial difference between photons and non-relativistic quantum mechanical point particles is that, while non-relativistic particles can move at any speed, photons are limited to travelling at only one speed, namely the speed of light $c$. As we shall see later in Section \ref{Sec:DynHam}, the dynamical Hamiltonian of a free particle in configuration space is given by
\begin{eqnarray} \label{However2}
H_{\rm dyn} &=& \hat v \, \hat p\, .
\end{eqnarray}
In the case of photons, this Hamiltonian simplifies to \cite{AMC}
\begin{eqnarray}
\label{photonHam}
H_{\rm dyn} &=& \sum_{s=\pm 1} s c \, \hat p \, . 
\end{eqnarray}
The summation over the direction of propagation $s$ is needed as photons may travel left or right along the $x$ axis. In addition, it has recently been shown that single photons also have a position representation and can be assigned a wave function \cite{Jake,Daniel}.

If an experiment can be modelled using both classical and quantum physics, both descriptions must yield the same predictions; otherwise, it would be possible to contradict quantum physics with classical experiments. Hence, we usually claim that classical mechanics and QM are consistent when the position and momentum expectation values, $\langle x(t) \rangle$ and $\langle p(t) \rangle$ coincide with $x(t)$ and $p(t)$ in the classical theory. This criterion became known as Ehrenfest's theorem \cite{Ehrenfest,greensite}. It has been noticed by several authors \cite{Wheeler,Messiah,Hall,Klauder,Dorje,Caslav}, however, that in basic QM this condition only holds in certain situations, such as free space, i.e.~in the absence of a potential $V(x)$. In particular, it was Ehrenfest who pointed out in Ref.~\cite{Ehrenfest} that consistency applies in general only when we have very narrow wave packets that remain localised in position. Another example where Ehrenfest's theorem holds is a point particle inside a harmonic oscillator.

By construction, QM in configuration space is consistent with classical mechanics in the following sense. One consequence of demanding that the most classical quantum states evolve as in classical mechanics is that the position and velocity expectation values $\langle x(t) \rangle$ and $\langle v(t) \rangle$ of superposition states $|\psi (t) \rangle$ evolve like the expectation values of a classical statistical mixture \cite{Oxford}. To show that this is indeed the case, we now consider a quantum mechanical point particle that has been prepared in a state of the general form 
\begin{eqnarray} \label{A1}
   |\psi(t) \rangle &=& \sum_{n=1}^N \sqrt{P_n} \, |x_n (t),v_n(t) \rangle \, ,
\end{eqnarray}
where the $P_n$ are probabilities that add up to one. Using this notation, the expectation values $\langle x(t) \rangle$ and $\langle v(t) \rangle$ can be expressed as
\begin{eqnarray} \label{A2}
    \langle x(t) \rangle = \sum_{n=1}^N P_n \, x_n (t) \, , ~~
    \langle v(t) \rangle = \sum_{n=1}^N P_n \, v_n(t)\, .
\end{eqnarray}
These expectation values coincide with the expectation values of a classical particle which moves with probability $P_n$ along trajectories $x_n(t)$. Hence the above equations imply that
\begin{eqnarray} \label{A3}
    \langle \dot x(t) \rangle = \sum_{n=1}^N P_n \, \dot x_n(t) \, , ~~
     \langle \dot v(t) \rangle = \sum_{n=1}^N P_n \, \dot v_n(t) \, ,
\end{eqnarray}
which shows that the quantum states $|\psi (t) \rangle$ and classical statistical mixtures with the same initial probability distributions evolve in the same way. Hence, QM in configuration space which we introduce here has many similarities with the mechanics of de Broglie and Bohm \cite{Bohm,Duer}.

 \begin{figure}[t]
    \centering
    \includegraphics[width= 0.99 \linewidth] {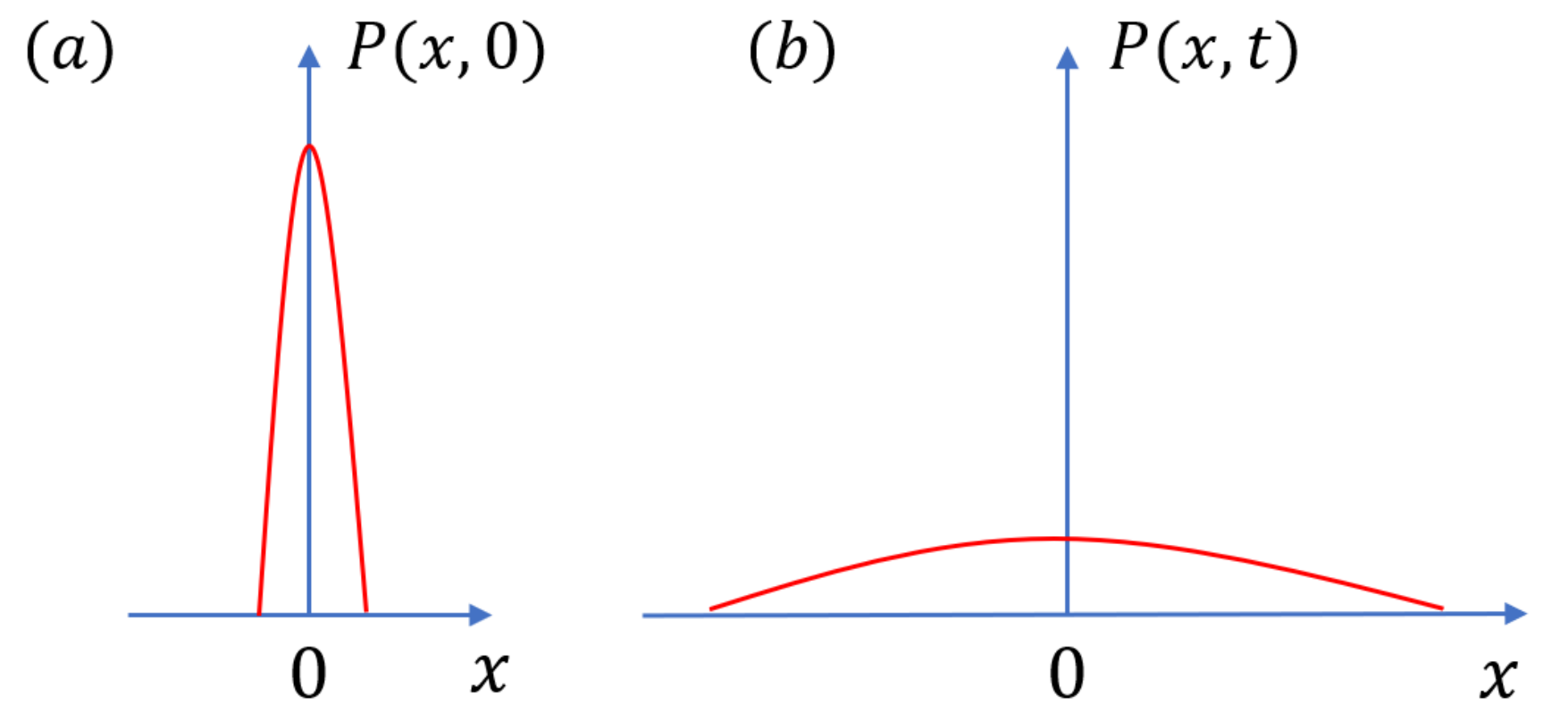}
    \caption{(a) The probability density $P(x,0)$ for finding a quantum mechanical point particle at position $x$ at an initial time $t=0$ when localised near $x=0$. (b) $P(x,t)$ for the same particle at a later time $t>0$ when the Hamiltonian of the particle is bound from below. As shown by Hegerfeldt in Refs.~\cite{Heger,Heger2}, basic QM implies that such a particle can be found anywhere along the $x$ axis with a non-zero probability density, even at coordinates $(x,t)$ outside the light cone of the particle at $t=0$. Hence, basic QM is not consistent with causality or with relativity.} \label{fig:b}
      \end{figure}

In basic QM, the probability $P(x,t)$ for finding a particle at a position $x$ at time $t$ will disperse over time, as illustrated in Fig.~\ref{fig:b}. What is more, for a particle prepared in an initially localised state $|x \rangle$, $P(x,t)$ is even non-zero at positions lying outside the relevant light cone, since $|x \rangle$ contains contributions from all possible momentum states $|p \rangle$. As shown by Hegerfeldt \cite{Heger,Heger2}, any system with a Hamiltonian that is bounded from below causes an initially localised particle to spread immediately everywhere. Although this might not constitute a problem in the non-relativistic regime \cite{Heger}, let us point out here that superluminal spreading does not necessarily arise in QM in configuration space, since its dynamical Hamiltonian $\hat H_{\rm dyn}$ has in general positive as well as negative eigenvalues. By construction, localised wave packets can remain localised when evolving in time. However, it is also possible to construct wave packets with infinite dispersion.

In addition to a position and a velocity observable, $\hat x$ and $\hat v$, QM in configuration space supports a momentum operator $\hat p$. Analogous to the canonical momentum in basic QM, we define the momentum operator $\hat p$ in the following as the generator for spatial translations. In this way, momentum expectation values are automatically conserved in situations with spatial translational invariance. As we shall see below, both operators $\hat v$ and $\hat p$ describe different quantities. As in basic QM, the momentum expectation value of a particle depends on the shape of its wave packet \cite{Griffith}. Measuring the momentum of a particle is less straightforward and requires, for example, an elastic collision with another particle. The same does not apply to velocity expectation values which tell us how much time it takes a particle to travel a certain distance. For example, Chang \cite{Chang 1,Chang 2,Chang 3} previously treated mass in the same way as momentum and energy to gain a better understanding why mass and energy can be converted into each other. The difference between momentum and velocity was also noticed by other authors \cite{krasno,krasno2}. 

To better illustrate the difference between the momentum and the velocity of a quantum mechanical point particle, let us have a closer look at the case of free fall. In this situation, a quantum mechanical point particle experiences a constant acceleration $g$ which is the same at any height, i.e.~independent of where we place our coordinate system. Spatial translational symmetry implies that the expectation values of the canonical momentum $\hat p$ must be conserved, since $\hat p$ is linked to the spatial derivative of the wave function. The same does not apply to the velocity expectation values of the particle which increase linearly in time, since the particle is accelerating. As we shall see below, this observation is in agreement with the predictions of QM in configuration space. However, basic QM predicts a momentum dynamics which is not compatible with spatial translational symmetry. For a discussion of some of the intricacies involved in defining momentum in classical mechanics see e.g.~Ref.~\cite{Goyal}. 

One prominent feature of quantum physics is Heisenberg's uncertainty principle which states that the position $\hat x$ and the momentum $\hat p$ of a quantum mechanical point particle cannot be known simultaneously to an arbitrarily high precision. The observables $\hat x$ and $\hat p$ are canonical variables and, in particular, 
\begin{eqnarray} \label{Sevi7}
    [ \hat x, \hat p ] &=& {\rm i} \hbar \, .
\end{eqnarray}
As we shall see below, this commutator relation also applies in configuration space as long as we identify the momentum operator $\hat p$ with the generator for spatial translations. Using this momentum definition, momentum expectation values are automatically conserved when the same physics applies everywhere. Moreover, it implies that $\hat x$ and $\hat p$ are related via a Fourier transform. In contrast to this, the position and the velocity of quantum mechanical point particles can be localised simultaneously, and we have
\begin{eqnarray} \label{Sevifinal}
    [ \hat x, \hat v ] &=& 0 \, .
\end{eqnarray}
In configuration space, velocity and momentum are represented by two very different commuting observables. A possible test of QM in configuration space is therefore to compare uncertainties of position, momentum and velocity measurements. While Heisenberg's uncertainty relations restrict the precision of momentum measurements, the same is no longer true for velocity measurements.
 
  \begin{figure}[t]
    \centering
    \includegraphics[width= 0.99 \linewidth] {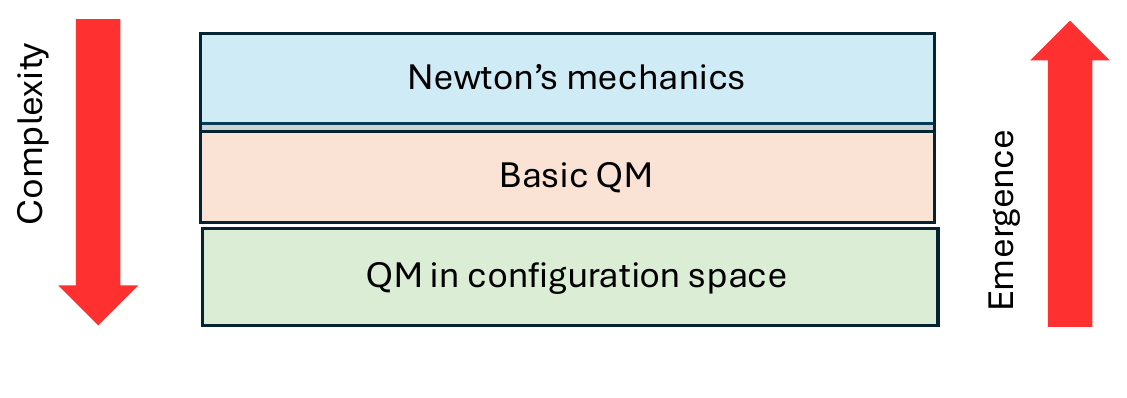}
    \caption{Physics often offers different theories for modelling the same situation. Those theories must be consistent with each other in the sense that less complex theories emerge from more complex ones, when certain approximations are applied. For example, as well shall see below, basic QM emerges from QM in configuration space when we restrict ourselves to a fixed mass $m$ and approximate the canonical momentum $\hat p$ by $m \, \hat v$. In addition, the Hamiltonian of the system needs to be adjusted (cf.~Eq.~(\ref{HHH})) such that expectation values evolve as expected. Classical physics emerges from QM in configuration space when its quantum states are replaced by classical statistical mixtures (cf.~Eq.~(\ref{A3})). The formalism which we present here is nevertheless non-classical; for example, simultaneous measurements are limited by uncertainty relations.} \label{fig:new}
      \end{figure}
 
Needless to say, over the last century QM has successfully modelled a wide range of experiments. An early example is the calculation of the emission spectrum of the hydrogen atom to a very high precision \cite{Bohr}. It is therefore important to show that basic QM emerges from QM in configuration space under certain approximations, as illustrated in Fig.~\ref{fig:new}. The main difference between basic QM and QM in configuration space is that the former lacks a velocity operator $\hat v$. As we shall see below, basic QM emerges from the formalism which we introduce here when we replace $\hat v$ with $\hat p/m$ where $m$ is fixed. Here the mass $m$ is an intrinsic property of the particle and can therefore be represented by a scalar. In addition, we need to replace the dynamical Hamiltonian $\hat H_{\rm dyn}$ with the standard Hamiltonian in basic QM \cite{Griffith}
\begin{eqnarray} \label{HHH}
\hat H_{\rm BQM} &=& {{\hat p}^2 \over 2 m} + V({\hat x})
\end{eqnarray}
of non-relativistic particles as this is the only way to ensure that the position and momentum expectation values $\langle x (t) \rangle $ and $\langle p (t) \rangle $ evolve according to Hamilton's equations of motion (cf.~Eq.~(\ref{Hamilton1})).

Despite considering different observables and different equations of motion, there are many similarities between basic QM and QM in configuration space. For example, as we shall see below, by construction both theories predict exactly the same physical trajectories of the position expectation values $\langle x(t) \rangle$ of mechanical point particles. Nevertheless, the formalism we introduce here offers clear advantages, avoiding some pitfalls highlighted in earlier approaches (cf.~e.g.~Refs.~\cite{Klauder,EUG} for more details). Notably, certain areas of quantum physics—such as quantum electrodynamics and condensed matter physics  \cite{Arovas}—often rely on phenomenological models to reconcile experimental results with theoretical  predictions. Indeed, previous successes of these models \cite{Doppler,AMC,Unruh,Casimir} have inspired the formalism we introduce here.

Another advantage of modelling quantum mechanical point particles in configuration space is that it offers a much wider range of possible initial states due to the increased size of its Hilbert space ${\cal H}$. It also provides more clarity regarding the physical meaning of its variables and the origin of its equations. Although further investigations are needed, we suspect that our approach might shine new light on situations which range from quantum transport problems to situations that we currently approach using non-Hermitian quantum physics \cite{Bender,Jake2}. Our theory also might provide new insights into quantum relativity \cite{QR} and into the emergence of chaotic behaviour in the quantum theory of mechanical systems due to its improved consistency between quantum and classical physics \cite{chaos}. 

The paper is organised as follows. In Section \ref{sec2}, we review the basic formalism of QM using Dirac notation, thereby also introducing the notation that we use throughout the remainder of the paper. We then construct wave functions and the basic observables of QM in configuration space in Section \ref{sec3}. In addition, it is shown here that momentum and position expectation values are still subject to Heisenberg's uncertainty relation. Afterwards, in Section \ref{sec4}, we take a closer look at the non-relativistic case and derive the dynamical Hamiltonian $\hat H_{\rm dyn}$ of quantum mechanical point particles in the presence of a non-zero potential $V(x)$. Moreover, we have a closer look at the emergence of the basic quantum mechanics from our configuration space formulation and at the implications of our approach for wave-particle duality. Finally, we review our findings in Section \ref{sec5}.

\section{The basic formulation of quantum mechanics} \label{sec2}

After the conceptual and mathematical foundations of quantum mechanics were laid about a century ago \cite{Dirac,Hi,Schr,Von}, quantum mechanics established itself as an independent branch of physics, and a large number of popular textbooks have been written about it \cite{Griffith,EUG,Englert,Rae,Franz,Liboff,MQM}. In this section, we review the main findings of  basic QM and introduce the notation that is used throughout the remainder of this paper. The state of a quantum mechanical point particle at time $t$ is represented by a complex state vector $|\psi(t) \rangle$ which belongs to a complex Hilbert space $\mathcal{H}$. For simplicity, we only study the one-dimensional case of a single particle that moves in the presence of a potential $V(x)$ along the $x$-axis. In the following, we consider both the position and momentum representations. Finally, we have a closer look at uncertainty relations.

\subsection{A single particle in position space} \label{21}

Wave-particle duality tells us that a quantum mechanical point particle can be at different positions $x$ at the same time. Its state vector $|\psi(t) \rangle$ is therefore, in general, a superposition of localised states $|x \rangle$. To ensure that a localised particle with a well-defined position $x$ cannot be found at a position $x \neq x'$, the states $|x \rangle$ and $|x' \rangle$ must be orthogonal to each other. In addition, they must be normalised. Hence, we require that 
\begin{eqnarray} \label{ortho}
\langle x|x' \rangle &=& \delta(x-x') \, ,
\end{eqnarray}
with the Dirac delta-function $\delta(x-x')$ given by
\begin{eqnarray}\label{deltax}
\delta(x-x') &=& \frac{1}{2\pi \hbar} \int_{\mathbb{R}} {\rm d}p \, {\rm e}^{{\rm i} p (x-x')/\hbar} \, .
\end{eqnarray}
Here ${\mathbb R} =(-\infty,\infty)$ denotes the real numbers. Since the states $|x \rangle$ form a complete orthonormal basis in the Hilbert space $\mathcal{H}$ of quantum mechanical point particles,
\begin{eqnarray} \label{identity}
\hat I &=&\int_{\mathbb R}{\rm d}x \, |x\rangle\langle x|
\end{eqnarray}
is the identity operator. Taking into account that $\hat I \, |\psi (t) \rangle= |\psi (t) \rangle$, we find that the state vector $|\psi (t) \rangle$ can be written as
\begin{eqnarray} \label{wave fun}
|\psi (t) \rangle &=& \int_{\mathbb R} {\rm d}x \, |x \rangle \, \psi(x,t) \, ,
\end{eqnarray}
with the complex wave function $\psi(x,t)$ given by $\psi(x,t) = \langle x|\psi(t) \rangle$. By definition, $|\psi(x,t)|^2 \, {\rm d}x$ represents the probability to find the particle at time $t$ between $x$ and $x + {\rm d}x$. Since the probability of finding the quantum mechanical point particle anywhere along that axis must equal one, we require that
\begin{eqnarray} 
\int_{\mathbb{R}} {\rm d}x \, |\psi(x,t) |^2 &=& 1 \, .
\end{eqnarray}
This condition applies when the state vector $|\psi(t) \rangle$ is normalised and $\langle \psi (t)|\psi (t) \rangle = 1$. 

We know that any physical observable $A$ of a quantum mechanical point particle can be represented by a Hermitian operator $\hat A$. By construction   
\begin{eqnarray} \label{observe}
\langle A(t) \rangle &=& \langle \psi(t) |\hat{A}|\psi (t) \rangle
\end{eqnarray}
coincides with the expectation value of $A$ at time $t$ when measured on systems prepared in the state $|\psi (t) \rangle$. The above equations can be used to show that 
\begin{eqnarray}\label{expx} 
\langle x(t) \rangle &=& \int_{\mathbb{R}} {\rm d}x \, x \, |\psi(x,t)|^2
\end{eqnarray}
is the expectation value of the position of a quantum mechanical point particle at time $t$. Hence its position operator $\hat x$ must equal
\begin{eqnarray} \label{expx2}
\hat{x} &=& \int_{\mathbb{R}} {\rm d}x \, |x \rangle \, x \, \langle x| \, .
\end{eqnarray}
When applied to the state vector $| \psi (t) \rangle$, this operator multiplies the wave function $\psi(x,t)$ with $x$ since
\begin{eqnarray}\label{16}
\hat x \, |\psi(t)\rangle&=& \int_{\mathbb{R}} {\rm d}x \, |x \rangle \, x \,\psi(x,t) \, .
\end{eqnarray}
Eq.~(\ref{expx2}) moreover shows that $|x \rangle$ is an eigenvector of $\hat x$ with $x$ being the corresponding eigenvalue. This is not surprising since the eigenvectors of any observable $\hat A$ are the states that reveal the corresponding eigenvalue upon measurement with maximum probability.

In the following, we assume that the momentum operator $\hat p$ of a quantum mechanical point particle is the generator of spatial translations. By defining momentum in this way, it is clear that momentum expectation values are conserved in situations that have spatial translational invariance \cite{Noether,Kosmann-Schwarzbach,AMC}.  As we will see at the end of this section when we discuss Weyl's commutator relation, this approach implies that 
\begin{eqnarray} \label{expx2p}
\hat{p} &=&  - {\rm i} \hbar \int_{\mathbb{R}} {\rm d}x \, |x \rangle \,\frac{\partial}{{\partial x}}\, \langle x| \,.
\end{eqnarray}
Consequently, applying $\hat p$ to the state vector $| \psi (t) \rangle$ replaces the wave function $\psi(x,t)$ with its spatial derivative and
\begin{eqnarray}\label{23}
\hat p \, |\psi(t)\rangle&=& -{\rm i}\hbar\int_{\mathbb{R}} {\rm d}x \, |x \rangle \, \frac{\partial}{\partial x} \,\psi(x,t) \, .
\end{eqnarray}
According to Eq.~(\ref{observe}), the momentum expectation value $\langle p(t) \rangle$ therefore equals $\langle \psi(t) |\hat p |\psi(t) \rangle$. 
The above momentum definition is analogous to the definition of energy. While momentum expectation values are conserved in situations with spatial translational symmetry, the conservation of energy expectation values occurs in situations with time translational invariance. 

\subsection{A single particle in momentum space} \label{sec2B}

One way of generating an alternative basis for the position states $|x \rangle$ is to relate them through a Fourier transform to the states $|p \rangle$ with $p \in \mathbb{R}$. More concretely, we assume in the following that 
\begin{eqnarray} \label{15bppp} 
|p\rangle &=& \frac{1}{\sqrt{2 \pi\hbar}} \int_{\mathbb{R}} {\rm d}x \, {\rm e}^{{\rm i} px/\hbar} \, |x \rangle \, .
\end{eqnarray}
Like the $|x \rangle$ states, the $|p\rangle$ states also form a complete orthonormal basis of the Hilbert space ${\cal H}$ of a quantum mechanical point particle since
\begin{eqnarray}  \label{deltap}
\langle p'|p \rangle &=& \delta(p-p') \, .
\end{eqnarray}
As the identity operator $\hat I$ can also be expressed as 
\begin{eqnarray} \label{Ip}
\hat I &=& \int_{\mathbb{R}} {\rm d}p \, |p\rangle\langle p| \, ,
\end{eqnarray}
the state vector $|\psi(t) \rangle$ can now be written as
\begin{eqnarray} \label{ortho2p}
| \psi (t) \rangle &=&\int_{\mathbb{R}}{\rm d}p \, \widetilde \psi(p,t) \, |p \rangle
\end{eqnarray}
with the complex wave function $\widetilde \psi(p,t) $ given by $\widetilde \psi(p,t) = \langle p|\psi(t) \rangle $. To identify the physical meaning of this new wave function, we now have a closer look at the momentum representation of $\hat p$.

Since the position states $|x \rangle$ are pairwise orthogonal and normalised, a closer look at Eq.~(\ref{15bppp}) shows that 
\begin{eqnarray} \label{overlap} 
\langle x|p\rangle &=& \frac{1}{\sqrt{2 \pi\hbar}} \, {\rm e}^{{\rm i} px/\hbar} 
\end{eqnarray}
while $\langle p|x \rangle = \langle x|p \rangle^* $.
Using the above equations and the fact that $\hat I \, \hat p \, \hat I = \hat p$, we can now show that 
\begin{eqnarray} \label{pink}
\hat{p}&=&\int_{\mathbb{R}} {\rm d}p \, |p \rangle \, p \, \langle p| \, .
\end{eqnarray}
This equation confirms that $p$ and $|p \rangle$ are the eigenvalues and the eigenstates of the momentum operator $\hat p$. In addition, the above equations can be used to show that the wave function $\widetilde \psi(p,t)$ relates to $\psi(x,t)$ via the inverse Fourier transform,
\begin{eqnarray}\label{fourier}
\widetilde \psi(p,t) &=& \frac{1}{\sqrt{2\pi\hbar}} \int_{\mathbb{R}}{\rm d}x \,  {\rm e}^{-ipx/\hbar} \, \psi(x,t)  \, .
\end{eqnarray}
Since Fourier transforms preserve the normalisation of wave functions, 
\begin{eqnarray}
\int_{\mathbb{R}} {\rm d}p \, |\widetilde \psi(p,t) |^2 &=& 1\, , 
\end{eqnarray}
and $\widetilde \psi(p,t)$ is also normalised. By definition, $ | \widetilde \psi(p,t) |^2 \, {\rm d}p$ is the probability for finding that the momentum of a quantum mechanical point particle with wave function $\widetilde \psi(p,t)$ lies between $p$ and $p + {\rm d} p$.

For completeness, we finally also have a look at the momentum representation of the position operator $\hat x$. Since the above equations can be used to show that 
\begin{eqnarray} \label{psix}
\langle p| \hat x| \psi(t) \rangle &=& {\rm i}\hbar \frac{\partial}{\partial p} \, \widetilde \psi(p,t) \, ,
\end{eqnarray}
we know that applying the position operator $\hat x$ in momentum space is equivalent to taking the $p$ derivative of the momentum space wave function $\widetilde \psi(p,t)$. Hence, the position operator $\hat x$ equals
\begin{eqnarray}\label{xink}
\hat x &=&{\rm i}\hbar \int_{\mathbb{R}}{\rm d}p\, |p\rangle \,\frac{\partial}{\partial p}\,\langle p| \, ,
\end{eqnarray}
in momentum space. 

\subsection{Uncertainty relations}\label{uncer}

A feature that highlights the fundamental difference between quantum and classical mechanics is Heisenberg's uncertainty principle. In quantum physics, any two physical observables $A$ and $B$ cannot be measured simultaneously unless their operators $\hat A$ and $\hat B$ commute. More concretely, Heisenberg's uncertainty relation suggests that
\begin{eqnarray}\label{un}
\Delta A \cdot \Delta B &\geq & \frac{1}{2} \, \left| \langle \psi (t)| \big [\hat A, \hat B \big] |\psi (t) \rangle \right | 
\end{eqnarray}
for simultaneous measurements of $A$ and $B$ on quantum states $|\psi (t) \rangle$. The uncertainties $(\Delta O)^2$ with $O=A,B$, which are defined as
\begin{eqnarray}
(\Delta O)^2 &=& \langle \psi (t)| \big( \hat O-\langle O(t) \rangle \big)^2 |\psi (t) \rangle \, ,
\end{eqnarray}
represent mean square deviations. Setting $\hat A=\hat x$ and $\hat B=\hat p$ and using Eqs.~(\ref{expx2}) and (\ref{expx2p}) to determine the commutator of the position and momentum operators $\hat x$ and $\hat p$, it is relatively easy to check that
\begin{eqnarray}\label{commuxp3}
    [\hat x, \hat p] &=& - {\rm i} \hbar\int_{\mathbb{R}} {\rm d}x \, |x \rangle \left ( x \,  \frac{\partial}{\partial x}- \frac{\partial}{\partial x}\, x \right) \, \langle x| \, , 
\end{eqnarray}
which leads to Eq.~(\ref{Sevi7}) after applying the product rule. Eq.~(\ref{un}) therefore yields the position-momentum uncertainty relation 
\begin{eqnarray} \label{S1}
\Delta x \cdot \Delta p &\geq& \frac{\hbar}{2} \, .
\end{eqnarray}
This equation implies that it is not possible to simultaneously measure both the position and the momentum of a quantum mechanical point particle with unlimited accuracy. In other words, a quantum mechanical point particle cannot be simultaneously localised in position and in momentum. However, it can be localised in position and velocity.

To illustrate more clearly that the momentum operator $\hat p$ in Eq.~(\ref{expx2p}) represents the generator for spatial translations, let us define the unitary translation operator $\hat{T}(\xi)$ in position space as
\begin{eqnarray} \label{K39}
\hat{T}(\xi) &=& {\rm e}^{-{\rm i} \xi \hat{p} / \hbar} 
\end{eqnarray}
where $\xi $ denotes a distance. This operator is also known as the displacement operator, since its action on a localised state $|x \rangle$ is given by
\begin{eqnarray}  \label{K40}
\hat{T}(\xi) |x\rangle &=& |x + \xi \rangle \, , 
\end{eqnarray} 
demonstrating that $\hat{T}(\xi)$ changes the position of quantum mechanical particles. Showing that this is the case can be done by applying the Taylor expansion to $\hat{T}(\xi)$ in Eq.~(\ref{K39}) before applying the operator to $|x \rangle$. Using Eq.~(\ref{K40}), it is relatively straightforward to verify Weyl's relation \cite{Weyl}
\begin{eqnarray} \label{K41}
{\rm e}^{-{\rm i} \xi \hat{p}/\hbar} \, {\rm e}^{-{\rm i} \mu \hat{x}/\hbar} 
&=& {\rm e}^{{\rm i} \xi \mu/\hbar} \, {\rm e}^{-{\rm i} \mu \hat{x}/\hbar} \, {\rm e}^{-{\rm i} \xi \hat{p}/\hbar} 
\end{eqnarray}
where $\mu$ is a constant. Both sides have the same effect on an initial state $|x \rangle$. Comparing both sides of the above equation while taking the Baker–Campbell–Hausdorff formula into account, we see that Weyl's commutator relation fully encapsulates the canonical commutation relation in Eq.~(\ref{Sevi7}).

\section{Quantum mechanics in configuration space} \label{sec3}

As we shall see below, when quantising classical point particles in a physically motivated manner, analogous to the quantisation of light in quantum electrodynamics \cite{Bennett,Jake,Daniel,AMC}, we obtain an alternative formulation of quantum mechanics, namely in configuration space. This section, closely follows the structure of Section \ref{sec2}. Again, we start by identifying the relevant Hilbert space of quantum mechanics by first having a closer look at the configuration space, i.e.~the fundamental space of all possible states, of classical mechanical point particles. Afterwards, we introduce a wave function for the particles as well as their position, momentum and velocity observables. In addition, we consider both the  position-velocity and momentum-velocity representations. As in basic QM, the position operator is diagonal in the position representation, while the momentum operator is diagonal in the momentum representation. We conclude this section by showing that our generalised description of quantum mechanical point particles does not violate Heisenberg's uncertainty relations despite the fact that the particle wave function $\psi(x,v,t)$ now also depends on the velocity $v$.

\subsection{A single particle in position-velocity space} \label{31}

Newtonian mechanics \cite{Goldstein,MQM} describes mechanical point particles by stating their position $x$ and their velocity $v$. For example, if we know the initial position $x(0)$ and the initial velocity $v(0)$ of a particle, we can predict its dynamics at all times $t$. As mentioned before, we therefore assume in the following that the positions and the velocities of a quantum mechanical point particle can be chosen independently. In addition, we promote the classical states $(x,v)$ of mechanical point particles
to distinguishable localised quantum states $|x,v\rangle$ with $x$ and $v$ specifying positions and velocities respectively. These states are the ``most classical" quantum states of QM in configuration space and form a basis in the Hilbert space ${\cal H}$ of quantum mechanical point particles.
This is in contrast to the position representation of basic QM, which we reviewed in Section \ref{21}. For example, we now consider a Hilbert space that is significantly larger than the one we had before. As we shall see below, the state $|\psi (t) \rangle$ of a quantum mechanical point particle at time $t$ is, in general, a superposition of all possible localised $|x,v \rangle$ states.

Since the $|x,v\rangle$ states correspond to distinguishable particle states, they must be pairwise orthogonal as stated in Eq.~(\ref{delta1}). Hence, the identity operator $\hat I$ acting on the extended Hilbert space ${\cal H}$ is now given by 
\begin{eqnarray} \label{delta2}
\hat I &=& \iint_{{\mathbb{R}}^2} {\rm d}x\, {\rm d}v \,|x,v\rangle \langle x,v| \, .
\end{eqnarray} 
Here the double integrals are taken over the parameter set ${\mathbb{R}}^2 = \{(x,v)\}$ with $x, v \in {\mathbb{R}}$. After defining the wave function $\psi(x,v,t)$ as $\psi(x,v,t) = \langle x,v| \psi(t)\rangle $, the state vector $|\psi(t)\rangle$ can be written as
\begin{eqnarray}\label{psittt}
|\psi(t)\rangle=  \iint_{{\mathbb{R}}^2} {\rm d}x\, {\rm d}v  \, {\psi(x,v,t)} \, |x,v\rangle \, .
\end{eqnarray}
The only difference to Eq.~(\ref{wave fun}) is that the basis vectors and the wave function now depend on the position $x$ and on the velocity $v$ of the quantum mechanical point particle. Since $| \psi(x,v,t) |^2$ represents the probability density for finding the particle at position $x$ and moving with velocity $v$ at time $t$, we can conclude that 
\begin{eqnarray}\label{iiii} 
\iint_{{\mathbb{R}}^2} {\rm d}x\, {\rm d}v  \,|\psi(x,v,t) |^2 &=& 1 \, .
\end{eqnarray}
This condition guarantees the normalisation of the state vector $|\psi(t) \rangle$.

As mentioned above, $|x,v \rangle$ describes a quantum mechanical point particle at position $x$ moving with exactly velocity $v$. Hence, the observables for the position and the velocity of the particle, $\hat x$ and $\hat v$, are given by 
\begin{eqnarray}\label{xop}
\hat{x} &=& \iint_{{\mathbb{R}}^2} {\rm d}x\, {\rm d}v \, |x,v\rangle \, x \, \langle x,v| \, , \notag \\
\hat {v} &=& \iint_{{\mathbb{R}}^2} {\rm d}x\, {\rm d}v  \, |x,v\rangle \, v \, \langle x,v| \, .
\end{eqnarray}
Hence, applying the position operator $\hat x$ and the velocity operator $\hat v$ to a state vector $|\psi(t) \rangle$ of a quantum mechanical point particle multiplies its wave function $ \psi(x,v,t)$ with $x$ and with $v$ respectively. Moreover, the above definitions ensure that the position and velocity expectation values $\langle x(t)\rangle $ and $\langle v(t)\rangle $ can be calculated using Eq.~(\ref{observe}).

As we shall see below, to construct the dynamical Hamiltonian $\hat H_{\rm dyn}$ of a quantum mechanical point particle in the presence of a potential $V(x)$, we moreover require operators that generate changes in position and in velocity. The first is the momentum operator $\hat p$ which is the generator of spatial translations and has many similarities with the momentum operator $\hat p$ in basic QM. More concretely, we define $\hat p$ in the following as 
\begin{eqnarray} \label{E85}
\hat p &=& -{\rm i}\hbar \iint_{{\mathbb{R}}^2} {\rm d}x \, {\rm d}v\,|x,v\rangle \, \frac{\partial}{\partial x} \, \langle x,v| \, ,
\end{eqnarray} 
in analogy to Eq.~(\ref{expx2p}). 
In addition, it is useful to also introduce an operator $\hat a$ which is the generator for changes in the velocity $v$ of a moving quantum mechanical point particle. In analogy to Eq.~(\ref{E85}), we define $\hat a$ as
\begin{eqnarray} \label{E85a}
\hat a &=& -{\rm i}\hbar \iint_{{\mathbb{R}}^2} {\rm d}x \, {\rm d}v \, |x,v\rangle \, \frac{\partial}{\partial v} \, \langle x,v| \, .
\end{eqnarray} 
The factor $ -{\rm i}\hbar $ has been added to ensure that $\hat a$ is a self-adjoint operator and that $\hat v$ and $\hat a$ satisfy the same commutation relation as $\hat x$ and $\hat p$. As we shall see in Section \ref{sec4}, in the presence of a force, $\hat a$ changes the velocity of a quantum mechanical point particle. We therefore refer to this operator in the following as the {\em acceleratum}. Analogously, as we shall see below, the momentum operator $\hat p$ changes the position of a particle with a non-zero velocity $v$.

\subsection{A single particle in momentum-velocity space} \label{32}

In the previous subsection, we replaced the position basis states $|x \rangle$ of basic QM with the new basis states $|x,v \rangle$. Analogously, we now replace the momentum basis states $|p \rangle$ of basic QM with the states $|p,v\rangle$ which describe quantum mechanical point particles with momentum $p$ moving with velocity $v$. Since we want the $|p,v\rangle$ states to have the same properties as the $|p \rangle $ states in basic QM, like diagonalising the momentum operator $\hat p$, we define them such that 
\begin{eqnarray} \label{xxkknn} 
\langle x,v|p,v'\rangle &=& \frac{1}{\sqrt{2 \pi\hbar}} \, {\rm e}^{{\rm i} px/\hbar} \,\delta(v-v') \, ,
\end{eqnarray}
in analogy to Eq.~(\ref{15bppp}) with $\langle p,v|x,v \rangle = \langle x,v|p,v\rangle^*$. Since all we have done is add the label $v$ to the position and momentum basis states, it is straightforward to check that many of the equations in Section \ref{sec2} still apply. For example, the basis states $|p,v \rangle$ form a complete orthonormal basis and 
\begin{eqnarray}  \label{deltak}
\langle p,v|p',v'\rangle &=& \delta(p-p') \delta(v-v') \, ,
\end{eqnarray}
in analogy to Eq.~(\ref{deltap}).

In the expanded Hilbert space ${\cal H}$, the identity operator $\hat I$ can now also be written as
\begin{eqnarray}\label{identitiyk}
\hat I &=& \iint_{{\mathbb{R}}^2} {\rm d}p \, {\rm d}v \, |p,v\rangle\langle p,v| \, .
\end{eqnarray}
In this representation, the state vector $|\psi(t) \rangle$ is defined as
\begin{eqnarray} \label{ortho2ppp}
| \psi (t) \rangle &=& \iint_{{\mathbb{R}}^2} {\rm d}p \, {\rm d}v \, \widetilde \psi(p,v,t) \, |p,v \rangle \, .
\end{eqnarray}
with 
\begin{eqnarray} \label{WFp}
\widetilde \psi(p,v,t) &=& \langle p,v|\psi(t) \rangle \, . 
\end{eqnarray}
This wave function is normalised when 
\begin{eqnarray}
\iint_{{\mathbb{R}}^2} {\rm d}p \, {\rm d}v \, |\widetilde \psi(p,v,t) |^2 &=& 1 \, .
\end{eqnarray}
 The wave function $\widetilde \psi(p,v,t) $ relates to the wave function $\psi(x,v,t) $ via the Fourier transform 
\begin{eqnarray}
\psi(x,v,t) &=&\frac{1}{\sqrt{2 \pi \hbar}} \int_{\mathbb{R}} {\rm d}p \, {\rm e}^{{\rm i} px/\hbar} \,\widetilde \psi(p,v,t) \, , 
\end{eqnarray}
with the inverse transformation given by the inverse Fourier transform.

As in the previous section, we find that the momentum operator $\hat p$ simplifies to 
\begin{eqnarray}
    \hat p &=&  \iint_{{\mathbb{R}}^2} {\rm d}p \, {\rm d}v \, |p,v\rangle\, p\,\langle p,v| \, ,
\end{eqnarray}
in the expanded momentum-velocity representation, in analogy to Eq.~(\ref{pink}). However, the acceleratum $\hat a$ in Eq.~(\ref{E85a}) does not simplify and equals
\begin{eqnarray} \label{E85b}
\hat a &=& -{\rm i}\hbar \iint_{{\mathbb{R}}^2} {\rm d}p \, {\rm d}v \, |p,v\rangle \, \frac{\partial}{\partial v} \, \langle p,v| \, , 
\end{eqnarray} 
in the basis of the $|p,v \rangle$ states. When transforming the position operator $\hat x$ in Eq.~(\ref{xop}) into the momentum-velocity representation, we find that it can also be written as
\begin{eqnarray}
    \hat x&=&{\rm i}\hbar \iint_{{\mathbb{R}}^2} {\rm d}p \, {\rm d}v \,|p,v\rangle \, \frac{\partial}{\partial p} \, \langle p,v| 
\end{eqnarray}
in analogy to Eq.~(\ref{xink}). Moreover, the velocity operator in Eq.~(\ref{xop}) becomes 
\begin{eqnarray}
    \hat v&=& \iint_{{\mathbb{R}}^2} {\rm d}p \, {\rm d}v \,|p,v\rangle \,v\,\langle p,v| \, . 
\end{eqnarray}
Before having a closer look at the dynamics of the quantum states and the expectation values of quantum mechanical point particles in Section \ref{sec4}, let us comment on the uncertainty relations for position, momentum, velocity, and acceleratum measurements.

\subsection{Uncertainty relations} \label{Uncertainties}

Given the formal similarities between the set of operators $\hat x$ and $\hat p$ and the set of operators $\hat v$ and $\hat a$, it is not surprising that the commutator relation between the position and the momentum operator and between the velocity and the acceleratum are the same. More concretely, one can now show that
\begin{eqnarray}\label{xpv}
    [\hat x, \hat p] = [\hat v, \hat a] &=& {\rm i} \hbar \, ,
\end{eqnarray}
in our expanded description of quantum mechanical point particles. All other operators commute. Using, for example, Eqs.~(\ref{xop})-(\ref{E85a}), we can confirm that  
\begin{eqnarray}\label{xvp}
     [\hat x, \hat v] = [\hat p, \hat a] = [\hat x, \hat a] = [\hat p, \hat v] &=& 0 \, . 
\end{eqnarray}
This means that we still have Heisenberg's uncertainty relations, but their interpretation has changed. The description which we present here allows us to simultaneously specify, for example, the position and the velocity expectation values $\langle x(t) \rangle$ and $\langle v(t) \rangle$ of a quantum mechanical point particle. Similarly, we can measure its momentum and its velocity expectation values $\langle p(t) \rangle$ and $\langle v(t) \rangle$ and so on simultaneously with arbitrarily high precision. However, using the commutator relations in Eq.~(\ref{xpv}), we again obtain the position-momentum uncertainty relation in Eq.~(\ref{S1}) for position and momentum measurements on an ensemble of quantum mechanical point particles which have all been prepared in the same state. Analogously, the commutator relation in Eq.~(\ref{xpv}) implies the velocity-acceleratum uncertainty relation
\begin{eqnarray} \label{S1zzz}
\Delta v \cdot \Delta a &\geq& \frac{\hbar}{2} \, .
\end{eqnarray}
This means that the position and the momentum are still canonical variables. In addition, we now have a new set of canonical variables for quantum mechanical point particles, namely $\hat{v}$ and $\hat{a}$.

To illustrate that the acceleratum $\hat a$ which we introduced in Eq.~(\ref{E85a}) represents the generator for velocity changes, we now proceed as in Section \ref{uncer} and define a unitary translation operator $\hat{R}(\alpha)$ in position space as
\begin{eqnarray} \label{K39b}
\hat{R}(\alpha) &=& {\rm e}^{-{\rm i} \alpha \hat{a} / \hbar} 
\end{eqnarray}
where $\alpha$ denotes a velocity. Using again a Taylor expansion, we now find that the action of this operator on a velocity eigenstate state $|x,v \rangle$ is given by
\begin{eqnarray}  \label{K40b}
\hat{R}(\alpha) \, |x,v\rangle &=& |x, v+\alpha \rangle \, ,
\end{eqnarray} 
in analogy to Eq.~(\ref{K40}). This equation demonstrates that $\hat{R}(\alpha) $ generates changes in the velocity without affecting the position of quantum mechanical particles. Moreover, by applying both sides of the equation below to a velocity eigenstate state $|x,v \rangle$ while using Eq.~(\ref{K40b}), one can check the validity of Weyl's relation \cite{Weyl}
\begin{eqnarray} \label{K41b}
{\rm e}^{-{\rm i} \alpha \hat{a}/\hbar} \, {\rm e}^{- {\rm i} \beta \hat{v}/\hbar} 
&=& {\rm e}^{{\rm i} \alpha \beta /\hbar} \, {\rm e}^{-{\rm i} \beta \hat{v}/\hbar} \, {\rm e}^{-{\rm i} \alpha \hat{a}/\hbar} \, ,
\end{eqnarray}
in analogy to Eq.~(\ref{K41}), where $\beta$ is a constant. Notice that this equation encapsulates the canonical commutation relation $ [\hat v, \hat a] = {\rm i} \hbar$ in Eq.~(\ref{xpv}).

\section{From consistency and emergence to wave-particle duality} \label{sec4}

In the previous section, we introduced the state vector $|\psi(t)\rangle$, the basic observables $\hat x$ and $\hat v$ and the operators $\hat p$ and $\hat a$ of a quantum mechanical point particle which moves along the $x$ axis. The main purpose of this section is to identify the equation of motion for $|\psi(t) \rangle$ in the presence of a potential $V(x)$. Let us assume that the time derivative of $|\psi(t) \rangle$ can, as usual, be specified by the Schr\"odinger equation of basic QM \cite{Griffith},
\begin{eqnarray} \label{newschh}
{\rm{i}}\hbar \,\frac{\rm d}{{\rm d} t} \, |\psi(t)\rangle  &=& \hat H_{\rm dyn} \, |\psi(t)\rangle\, , 
\end{eqnarray}
where $\hat H_{\rm dyn}$ denotes the dynamical Hamiltonian of the particle. An equation of this form should apply since the dynamics of physical objects with no memory effects depend only on their current state $|\psi(t)\rangle$ and its fully determined by its time derivative. As Eq.~(\ref{newschh}) shows, the dynamical Hamiltonian $\hat H_{\rm dyn} $ is by definition the generator for temporal translations. As we shall see in the following, it is therefore also closely related to the energy of the quantum mechanical point particle \cite{Noether,Kosmann-Schwarzbach}.

In Section \ref{Sec:DynHam}, we first have a closer look at the dynamics of the ``most classical" quantum states $|x,v \rangle$.  As discussed in the Introduction, these states have a well-defined position $x$ and a well-defined velocity $v$ and can be used to describe classical experiments. Hence, they must remain localised in $x$ and in $v$ and their parameters $x(t)$ and $v(t)$ must evolve as in classical mechanics, namely according to Eq.~(\ref{Newton4}). This observation allows us to identify the dynamical Hamiltonian $\hat H_{\rm dyn}$ of quantum mechanical point particles in Eq.~(\ref{newschh}). Moreover, since the $|x,v \rangle$ states form a complete basis in the extended Hilbert space of quantum mechanical point particles, we also automatically know how general quantum states, i.e.~superpositions of the $|x,v \rangle$ basis states, evolve in time. Afterward, in Section \ref{sec42}, we show that the basic formulation of QM is essentially a semi-classical approximation of QM in configuration space. Finally, in Section \ref{sec43}, we revisit wave–particle duality and highlight that it is enhanced in the description that we present here.

\subsection{The dynamical Hamiltonian}
\label{Sec:DynHam}

Suppose a quantum mechanical point particle has been prepared in an initial state $|\psi(0) \rangle$ with a known wave function $\psi(x,v,0)$. Without restrictions, the time derivative of the wave function $\psi(x,v,t)$ of the particle at a later time $t$ can be shown to be of the general form
\begin{eqnarray}\label{difft}
{{\rm d} \over {\rm d} t} \psi(x,v,t) &=& \left[ \frac{\partial x}{ \partial t} \, \frac{\partial}{ \partial x} + \frac{\partial v}{ \partial t} \, \frac{\partial}{ \partial v} + \frac{\partial}{ \partial t}  \right] \psi(x,v,t) \, . ~~~
\end{eqnarray}
Since the $|x,v \rangle$ states evolve as in classical physics, the partial derivatives $\partial x/\partial t$ and $\partial v/\partial t$ must be the same as in Eq.~(\ref{Newton4}). Moreover, the fact that probability amplitudes move around configuration space, but do not change in time, implies that
\begin{eqnarray}\label{difft122}
\frac{\partial}{\partial t} \psi(x,v,t) &\equiv & 0 \, .
\end{eqnarray}
This equation ensures that the probability for a particle to have the coordinates $(x,v)$ at time $t$ is the same as the probability  to have the coordinates $(x+ v \Delta t,v+ f(x) \Delta t)$ at time $t+ \Delta t$. It also removes a constant term from the Hamiltonian of the QM system, which has no physical consequences. Consequently, we find that
\begin{eqnarray}\label{difft3}
{{\rm d} \over {\rm d} t} \psi(x,v,t) &=& \left[ v \, \frac{\partial}{ \partial x} + f(x) \, \frac{\partial}{ \partial v} \right] \psi(x,v,t) \, .
\end{eqnarray}
Eqs.~(\ref{xop})-(\ref{E85a}) therefore suggest that the dynamical Hamiltonian $\hat H_{\rm dyn}$ must be given by
\begin{eqnarray} \label{However}
   \hat H_{\rm dyn}&=& \hat{v} \, \hat{p}+f(\hat{x}) \, \hat{a} \, .
\end{eqnarray}
One can check that this is indeed the case by substituting Eq.~(\ref{However}) into the Schr\"odinger equation (\ref{newschh}) while considering the position-velocity representation of state vectors. The above Hamiltonian is automatically self-adjoint, since $\hat v$ and $\hat p$ as well as $\hat x$ and $\hat a$ commute. It is worth mentioning that both terms in the dynamical Hamiltonian $\hat H_{\rm dyn}$ have units of energy. Moreover, Eq.~(\ref{However}) simplifies to Eq.~(\ref{However2}) in free space when the potential $V(x) = 0$.

Notice that the dynamical Hamiltonian $\hat H_{\rm dyn}$ in Eq.~(\ref{However}) differs from the energy observables $\hat H_{\rm class}$ of quantum mechanical point particles that are obtained when applying the correspondence principle to the expression for the energy of a point particle in classical mechanics. More concretely, the correspondence principle suggests that $\hat H_{\rm class}$ equals  
\begin{eqnarray}\label{HHH1}
\hat H_{\rm class} &=& {m \hat v^2 \over 2} + V(\hat x)
\end{eqnarray}
where $m$ denotes a constant parameter, namely the respective mass of the particle. Fortunately, using Eq.~(\ref{newschh}), one can show that
\begin{eqnarray} \label{However3}
\left[ \hat H_{\rm class}, \hat H_{\rm dyn} \right]&=&0 \, .
\end{eqnarray}
Hence, the expectation values of the above energy observable are always conserved, and QM in configuration space does not contradict classical mechanics.  Finally, let us point out that the expectation values of $\hat H_{\rm class}$ would not be conserved if we were to replace the observable $m \hat v$ in Eq.~(\ref{HHH1})  with $\hat p$. Moreover, we do not yet know the physical meaning of the expectation values of the terms that appear in the dynamical Hamiltonian, namely $ \hat{v} \, \hat{p}$ and $f(\hat{x}) \, \hat{a}$, which might explain why we usually do not consider these quantities in classical mechanics. 

Now that the Hamiltonian $\hat H_{\rm dyn}$ has been identified, we can predict the dynamics of arbitrary initial states and analyse the dynamics of expectation values of physical observables $\hat A$. Using Schr\"odinger's equation, one can show that
\begin{eqnarray} \label{td}
\frac{\rm d}{{\rm d}t}\langle A(t) \rangle &=& - \frac{\rm i}{\hbar} \, \left \langle \left[ \hat{A},\hat{H}_{\rm{dyn}} \right] \right \rangle \, .
\end{eqnarray}
Using Eqs.~(\ref{However}) and (\ref{xpv})-(\ref{xvp}), we therefore find that 
\begin{eqnarray} \label{td6}
\frac{\rm d}{{\rm d}t}\langle x(t) \rangle = \langle v \rangle \, , ~~
\frac{\rm d}{{\rm d}t}\langle v(t) \rangle = \langle f(x) \rangle 
\end{eqnarray}
in agreement with Eq.~(\ref{Newton4}). For example, when $\langle f(x) \rangle = f(\langle x \rangle) $, these two equations form a closed set of differential equations which means that we only need to consider position and velocity to predict the trajectory of a quantum particle in configuration space for any given initial state $|\psi(0) \rangle$. However, in general, we also need to have a closer look at the dynamics of the expectation values of the momentum $\hat p$ and the acceleratum $\hat a$ of the particle. Using again the above equation, one can moreover show that   
\begin{eqnarray} \label{td2}
\frac{\rm d}{{\rm d}t}\langle p(t) \rangle = - \left \langle {{\rm d} f(x) \over {\rm d}x} \, a \right \rangle \, , ~~
\frac{\rm d}{{\rm d}t}\langle a(t) \rangle = - \langle p \rangle \, . 
\end{eqnarray}
To show that this is the case, we write $f(x)$ without restrictions as a Taylor expansion,
\begin{eqnarray} \label{td3}
f(\hat x) &=& \sum_{n=0}^\infty c_n \, \hat x^n 
\end{eqnarray}
where the $c_n$ are real coefficients, and notice that
\begin{eqnarray} \label{td4}
\left[ f(\hat x), \hat p \right] &=& \sum_{n=0}^\infty c_n \left[ \hat x^n , \hat p \right] 
= - \sum_{n=0}^\infty n \, c_n \, \hat x^{n-1} \, .
\end{eqnarray}
For example, in free space, we have $V(x) = f(x) = 0$ and hence one can easily check that both the velocity and the momentum of the particle are conserved in time. Next we consider a mechanical point particle in free fall. In this case, $V(x) = - mgx$ where $x$ denotes the initial height of the particle and where $g$ denotes the gravity acceleration constant. Consequently, $f(x) = g$ which implies that
\begin{eqnarray} \label{td5}
\frac{\rm d}{{\rm d}t}\langle v(t) \rangle = g \, , ~~
\frac{\rm d}{{\rm d}t}\langle p(t) \rangle = 0 \, . 
\end{eqnarray}
As noted previously in the Introduction, the velocity of the particle increases linearly in time while its momentum is conserved due to the spatial translational invariance of the situation. Independent of where we place the coordinate system, the particle experiences the same physics, i.e.~the same acceleration $g$.

\subsection{The emergence of basic quantum mechanics} \label{sec42}

\begin{figure}[t]
\centering
    \includegraphics[width=0.7 \linewidth] {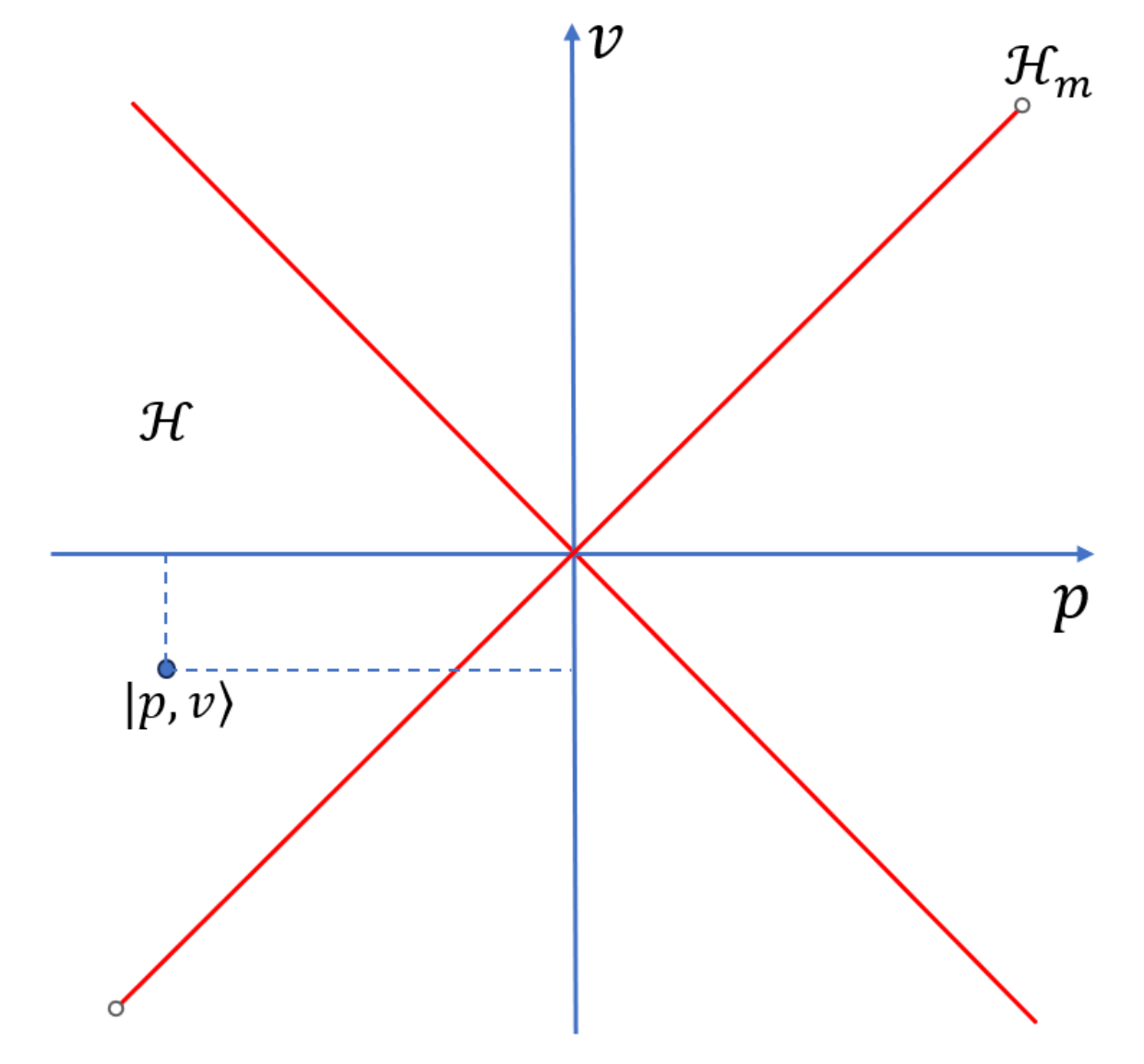}
    \caption{The momentum representation of the Hilbert space ${\cal H}$ of quantum mechanical point particles. In QM in configuration space, the state vector $|\psi (t) \rangle$ is in general a superposition of all basis states $|p,v \rangle$. These can each be represented by a point in the diagram. In contrast to this, basic QM restricts the states of a quantum mechanical point particle of a given mass $m$ effectively to $|p,v \rangle$ states with $|p/v|= m$, thereby reducing the total Hilbert space ${\cal H}$ to a subspace ${\cal H}_m$. The subspace ${\cal H}_m$ of particles of equal mass $m$ can be represented by two lines through the origin of the $(p,v)$-space diagram (red). As in classical physics, the mass $m$ is a parameter which identifies the type of particle that is being considered. When allowing for particles of any mass $m$, the Hilbert spaces of QM in configuration space and basic QM become the same.}
    \label{fig:a}
      \end{figure}

The basic formulation of QM has already been used successfully to explain experiments in many areas of physics, from quantum optics to condensed matter physics. It is therefore important to show that basic QM emerges from QM in configuration space, as illustrated in Fig.~\ref{fig:new}, when certain approximations are applied. As we shall see below, basic QM is essentially a semi-classical approximation of QM in configuration space. As we have seen in Section \ref{sec2}, basic QM only considers one set of canonical variables, $\hat x$ and $\hat p$. To reduce the four observables $\hat x$, $\hat v$, $\hat p$ and $\hat a$ of QM in configuration space to $\hat x$ and $\hat p$ requires three simplifications:
\begin{enumerate}
\item Since the Hilbert space of the basic formulation of QM is much smaller than the Hilbert space $\mathcal{H}$ of QM in configuration space, we decompose $\mathcal{H}$ as shown in Fig.~\ref{fig:a} into subspaces $\mathcal{H}_m$. Each of these subspaces contains all $\lvert p, v \rangle$ states satisfying $\lvert p/v \rvert = m$ as illustrated by the red lines in the diagram. In basic QM, a quantum mechanical point particle of a given mass $m$ can only occupy superposition states within $\mathcal{H}_m$. This means, $p$ and $v$ cannot be chosen independently and every particles has a well-defined mass $m$ which is a real parameter. 
\item We need to replace the velocity operator $\hat v$ of QM in configuration space such that
\begin{eqnarray}\label{F2}
\hat v &=& \hat p / m \, ,
\end{eqnarray}
where $\hat p$ is the canonical momentum operator and $m$ denotes the mass of the respective particle. In addition, we need to discard $\hat a$. There is no acceleratum in basic QM.
\item The main observables of the basic framework of QM are the position $x$ and the momentum $p$ with the intrinsic assumption that $\hat p=m \hat v$, as stated in Eq.~(\ref{F2}). In other words, basic QM needs to ensure that 
\begin{eqnarray} \label{td6extra}
\frac{\rm d}{{\rm d}t}\langle x(t) \rangle = \langle v \rangle \, , ~~
\frac{\rm d}{{\rm d}t}\langle p(t) \rangle = \langle m f(x) \rangle 
\end{eqnarray}
in analogy to Eq.~(\ref{td6}) in order to be in agreement with Eq.~(\ref{Newton4}). Since the commutators $[\hat x,\hat v]$ and $[\hat x,\hat p]$, however, are not the same (cf.~Eqs.~(\ref{Sevi7}) and (\ref{Sevifinal})), we also need to replace the dynamical Hamiltonian $\hat H_{\rm dyn}$ in Eq.~(\ref{However}) with the standard Hamiltonian $\hat H_{\rm BQM} $ in Eq.~(\ref{HHH}), while retaining the Schr\"odinger equation, i.e.~Eq.~(\ref{newschh}). 
\end{enumerate}
As we have discussed in detail in Section \ref{sec3}, in QM in configuration space, the momentum and the velocity of a quantum mechanical point particle correspond to commuting observables $\hat p$ and $\hat v$ but nevertheless describe two different physical properties. To measure the velocity expectation value $\langle v(t) \rangle$ of a particle, we need to measure, for example, the distance that it travels in a certain time interval. In contrast to this, the momentum expectation value $\langle p(t) \rangle$ can be measured, for example, by colliding the particle with another object while observing momentum conservation. The independence of the momentum and the velocity of quantum mechanical point particles is well reflected in QM in configuration space by the fact that the momentum depends only on the $x$ distribution of the wave function $\psi (x,v,t)$, but not on its velocity distribution. In contrast to this, the velocity of a quantum mechanical point particle does not depend on its position distribution.

\subsection{Wave-particle duality} \label{sec43}

As indicated by the title of this paper, the main motivation for the introduction of QM in configuration space is to fully realise the duality between the trajectories of classical point particles and the evolution of QM wave functions, which is not fully realised in basic quantum physics. For example, a photonic wave packet can be of any shape. Nevertheless, the photon only travels at one speed, namely the speed of light $c$. In contrast to this, the motion of quantum mechanical point particles in basic QM depends strongly on the shape of their wave packets. To illustrate the duality of particles and waves brought out in QM in configuration space, we show in the following that QM in configuration space also applies to photons. Although photons are carriers of electromagnetic fields, we ignore the presence of these field observables in the following discussion. A detailed account of how to obtain the electric and magnetic field observables of light through physically-motivated field quantisation schemes, and how to link those to monochromatic waves and local photon excitations, can be found in Refs.~\cite{Bennett,Jake,Daniel,AMC}.

For simplicity, we consider a single photon of a given polarisation $\lambda$ which travels along the $x$ axis in the $s$ direction, with $s = \pm 1$. As already mentioned above, photons move at only one speed, the speed of light $c$. For the mechanical point particles, the basic building blocks of the system are localised excitations prepared in states $|x,v \rangle$ with a well-defined position $x$ and velocity $v$. Analogously, the basis states of the wave packets of single photons with polarisation $\lambda$ which travel in the $s$ direction along the $x$ axis are states $|x,sc \rangle$ with
\begin{eqnarray}\label{Z1}
   |x,sc \rangle &=& a^\dagger_{s \lambda}(x) |0 \rangle 
\end{eqnarray}
with $|0 \rangle$ denoting the vacuum state of the electromagnetic field. Moreover, $a_{s \lambda}(x)$ and $a^\dagger_{s \lambda}(x)$ are locally-acting bosonic single-photon annihilation and creation operators that were introduced in Refs.~\cite{Jake,Daniel}. Alternatively, we can decompose the wave packets of single photons into monochromatic field excitations. In this case, their basis states $|p,sc \rangle$ are given by
\begin{eqnarray}\label{Z2}
   |p,sc \rangle &=& \tilde a^\dagger_{s \lambda}(k) |0 \rangle 
\end{eqnarray}
with $p = \hbar k$. Here $\tilde a_{s \lambda}(k)$ and $\tilde a^\dagger_{s \lambda}(k)$ are the annihilation and creation operators of monochromatic photons which relate to the $a_{s \lambda}(x)$ and the $a^\dagger_{s \lambda}(x)$ operators via a Fourier transform. This means that the quantum states of single photons occupy the same Hilbert space as quantum mechanical point particles in QM in configuration space, but with their velocities $v$ restricted to $v = \pm c$.

In free space, the photon does not experience any potentials, and $V(x)=0$. As shown in Refs.~\cite{Jake,Daniel,AMC}, in this case the dynamics of light of polarisation $\lambda$ can be described by the dynamical Hamiltonian 
\begin{eqnarray}\label{Z3}
   \hat H_{\rm dyn}&=& -{\rm{i}} \hbar c \sum_{s=\pm 1} \int_{\mathbb{R}} {\rm d}x \, a^\dagger_{s\lambda}(x) \, \frac{\partial}{ \partial x} \, a_{s\lambda}(x) \, .
\end{eqnarray}
When restricting the quantum state of the electromagnetic field to a single photon, this Hamiltonian simplifies to
\begin{eqnarray}\label{Z4}
   \hat H_{\rm dyn}&=&  -{\rm{i}} \hbar c \sum_{s=\pm 1} \int_{\mathbb{R}} {\rm d}x \, |x,sc \rangle  \langle 0| \, \frac{\partial}{ \partial x} \, |0 \rangle \langle x,sc| \, , 
\end{eqnarray}
since $a_{s\lambda}(x) = |0 \rangle \langle x,sc|$ in this case. One can then easily check that
\begin{eqnarray}\label{Z5}
   \hat H_{\rm dyn}&=&  -{\rm{i}} \hbar c \sum_{s=\pm 1} \int_{\mathbb{R}} {\rm d}x \, |x,sc \rangle \, \frac{\partial}{ \partial x} \, \langle x,sc| 
\end{eqnarray}
if the vacuum state is normalised and $\langle 0|0 \rangle = 1$. Comparing this operator with the momentum operator $\hat p$ in Eq.~(\ref{E85}), we see that the dynamical Hamiltonian $\hat H_{\rm dyn} $ of light can indeed be written as in Eq.~(\ref{photonHam}) in the Introduction. In principle, single photons can also experience non-zero potentials $V(x)$ and change their velocity. This requires the presence of an inhomogeneous dielectric medium with a changing refractive index $n(x)$, which can produce novel optical effects and is still a topic of ongoing research \cite{AMC,Pendry}. In conclusion, we can use QM in configuration space to describe the dynamics of single photons, as implied by wave-particle duality.

\section{Conclusions} \label{sec5}

There is no unique way of defining the momentum $p$ of a particle, even in classical physics \cite{Goyal}. For example, the momentum of a mechanical point particle could be defined as $p=mv$, where $m$ and $v$ denote its mass and its velocity, respectively. Another possible definition of the momentum $p$ is to demand that it is a conserved quantity for systems with spatial translational symmetry \cite{Noether}. In free space, where the potential $V(x) = 0$, both definitions are consistent with each other. Consider, however, the situation of a particle in free fall which experiences a constant acceleration. In this case, the same physics applies everywhere, which implies spatial translational invariance and hence momentum conservation. Unfortunately, this is not the case when we assume that $p=mv$, since the velocity of the particle and therefore also its momentum increase linearly in time. One reason why the definition of the momentum of particles is elusive, is that it does not make an appearance in the dynamical equations of mechanical particles, cf.~e.g.~Eq.~(\ref{Newton4}) but also Eq.~(\ref{td6}). In fact, the basic formalism of QM confuses momentum and velocity while imposing mass as a particle property (cf.~Eq.~(\ref{td6extra}). However, quantum physics offers a way of defining the momentum of particles in a unique way \cite{AMC}.

In this paper, we therefore quantise mechanical point particles in a physically-motivated way. To do so, we first identify the variables that are needed to characterise the dynamics of particles in classical mechanics, namely their position $x$ and their velocity $v$. Afterwards, we promote all possible distinguishable classical states $(x,v)$ to pairwise orthonormal quantum states $|x , v\rangle$. In this way, any classical experiment can be mapped onto a quantum state. To ensure that the quantum dynamics of these states are consistent with the dynamics of the associated classical system, we demand that the parameters $x$ and $v$ of the ``most classical" quantum states $|x , v \rangle$ evolve according to Newton's equations of motion (cf.~Eq.~(\ref{Newton4})) and choose the dynamical Hamiltonian $\hat H_{\rm dyn}$ accordingly. When preparing a quantum system in a superposition of $|x , v\rangle$ states, i.e.~in a general quantum state $|\psi \rangle$, expectation values can be shown to evolve in the same way as the expectation values of a classical particle prepared in $(x,v)$ with its probability distribution given by $| \langle x,v|\psi \rangle |^2$ (cf.~Eqs.~(\ref{A1})-(\ref{A3})). Nevertheless, QM in configuration space is not a classical hidden variable theory \cite{Bohm,Duer}, but a quantum theory that contains basic QM as a semi-classical approximation.

In QM in configuration space, a quantum mechanical point particle has a position and momentum operator, $\hat x$ and $\hat p$. In addition, we obtain a velocity operator $\hat v$ and an acceleratum $\hat a$. Having a closer look at the dynamical Hamiltonian $\hat H_{\rm dyn}$ in Eq.~(\ref{However}), we see that the momentum $\hat p$ generates spatial translations whenever states with non-zero velocity $\hat v$ are populated. By construction, its expectation values are conserved in case of spatial translational symmetry. Analogously, the acceleratum $\hat a$ induces changes of velocity for states which experience a force. In good agreement with Heisenberg's uncertainty relation, we find that $\hat x$ and $\hat p$ do not commute. The same applies to $\hat v$ and $\hat a$. As one can see from Eqs.~(\ref{xpv}) and (\ref{xvp}), the momentum and the velocity observables $\hat p$ and $\hat v$ of QM in configuration space commute and represent independent physical properties. Since the formalism of basic QM only considers position and momentum observables, it is essentially a semi-classical approximation of QM in configuration space. 

As mentioned already above, momentum expectation values are conserved in all situations with {\em spatial} translational invariance. Similarly, energy expectation values should be conserved in all situations with {\em time} translational invariance. As one can see from Eq.~(\ref{newschh}), the dynamical Hamiltonian $\hat H_{\rm dyn}$ is the generator for temporal translations. This suggests that its expectation values of $\hat H_{\rm dyn}$ are conserved when the potential $V(x)$ remains constant in time \cite{Noether}. However, the dynamical Hamiltonian has positive and negative eigenvalues \cite{Jake,Daniel}, while the the energy of particles should be always positive. Suppose
\begin{eqnarray}\label{Z6}
   \hat H_{\rm dyn} &=& \sum_{n=0}^\infty \int_{\mathbb{R}} {\rm d} E_n \, |E_n \rangle \, E_n \, \langle E_n |
\end{eqnarray}
represents the dynamical Hamiltonian in diagonal form. Then the energy observable of quantum mechanical point particles can be defined as \cite{AMC}
\begin{eqnarray}
	\label{Z7}
   \hat H_{\rm eng} &=& \sum_{n=0}^\infty \int_{\mathbb{R}} {\rm d} E_n \, |E_n \rangle \, |E_n| \, \langle E_n | \, .
\end{eqnarray}
In this way, energy is always positive but it is also conserved, since $\hat H_{\rm dyn} $ in Eq.~(\ref{Z6}) and $\hat H_{\rm eng} $ in Eq.~(\ref{Z7}) commute by construction. Classical mechanics, however, suggests a different definition of the energy observable, namely the operator $\hat H_{\rm class} $ in Eq.~(\ref{HHH1})). Fortunately, this does not create any problems, since $\hat H_{\rm dyn}$ and $\hat H_{\rm class} $ commute as well (cf.~Eq.~(\ref{However3})).  The predictions of QM in configuration space are therefore always consistent with the predictions of classical mechanics.

In addition, the formalism of QM in configuration space enhances wave-particle duality in comparison to the formalism of basic QM. In fact, this paper treats quantum mechanical point particles in exactly the same way as we treat photons in quantum electrodynamics \cite{Jake,Daniel,AMC}. The only difference between both types of particles is that photons can travel only at one speed, namely the speed of light $c$, while quantum mechanical point particles can move at any speed. This is important, since wave-particle duality is the cornerstone of quantum physics. In addition, we expect that QM in configuration space has applications in the modelling of chaotic systems and the modelling of quantum systems with non-Hermitian Hamiltonians with complex eigenvalues \cite{Bender}. Other areas of quantum physics, like Condensed Matter, already routinely employ phenomenological Hamiltonians, such as the Hubbard Hamiltonian \cite{Arovas}, with positive and negative eigenvalues in order to match experimental findings. Our hope is that applying QM in configuration space to such situations will help us gain a better understanding of complex quantum systems and will offer new tools for designing new devices with potential applications in quantum technology. In addition, we might gain new insight into relativistic quantum information \cite{Doppler,Unruh}. 
\\[0.5cm]
{\em Acknowledgments.} We thank Max Davies for the valuable discussions. A.B. thanks the Saudi Government Ministry of Education represented by the Saudi Arabian Cultural Bureau in London (SACB) for financial support. D.H. acknowledges financial support from the UK Engineering and Physical Sciences Research Council EPSRC (Grant No. EP/W524372/1).

\end{document}